\pdfoutput=1

\documentclass[reprint, amsmath, amssymb, aps, superscriptaddress]{revtex4-1}

\usepackage{graphicx}
\usepackage{dcolumn}
\usepackage{bm}
\usepackage{hyperref}
\usepackage{color}
\usepackage{lmodern}
\usepackage{epstopdf}

\begin{document}

\title{Thermal and magnetoelastic properties of $\alpha$-RuCl$_{3}$ in the field-induced low temperature states}

\author{Rico Sch{\"o}nemann}
 \email{rschoenemann@lanl.gov}
 \affiliation{MPA-MAGLAB, Los Alamos National Laboratory, Los Alamos, New Mexico 87545, USA.}
\author{Shusaku Imajo}
 \affiliation{The Institute for Solid State Physics, University of Tokyo, Kashiwa, Chiba 277-8581, Japan}
\author{Franziska Weickert}
 \affiliation{MPA-MAGLAB, Los Alamos National Laboratory, Los Alamos, New Mexico 87545, USA.}
\author{Jiaqiang Yan}
 \affiliation{Materials Science and Technology Division, Oak Ridge National Laboratory, Oak Ridge, Tennessee 37996, USA.}
 \affiliation{Department of Materials Science and Engineering, University of Tennessee, Knoxville, Tennessee 37996, USA.}
\author{David G. Mandrus}
 \affiliation{Materials Science and Technology Division, Oak Ridge National Laboratory, Oak Ridge, Tennessee 37996, USA.}
 \affiliation{Department of Materials Science and Engineering, University of Tennessee, Knoxville, Tennessee 37996, USA.}
\author{Yasumasa Takano}
 \affiliation{Department of Physics, University of Florida, Gainesville, Florida 32611, USA}
\author{Eric L. Brosha}
 \affiliation{MPA-11, Los Alamos National Laboratory, Los Alamos, New Mexico 87545, USA}
\author{Priscila F. S. Rosa}
 \affiliation{MPA-CMMS, Los Alamos National Laboratory, Los Alamos, New Mexico 87545, USA}
\author{Stephen E. Nagler}
 \affiliation{Neutron Scattering Division, Oak Ridge National Laboratory, Oak Ridge, Tennessee 37831, USA}
\author{Koichi Kindo}
 \affiliation{The Institute for Solid State Physics, University of Tokyo, Kashiwa, Chiba 277-8581, Japan}
\author{Marcelo Jaime}
 \email{mjaime@lanl.gov}
 \altaffiliation[present address: ]{Physikalisch-Technische Bundesanstalt, Braunschweig 38116, Germany}
 \affiliation{MPA-MAGLAB, Los Alamos National Laboratory, Los Alamos, New Mexico 87545, USA}

\date{\today}

\begin{abstract}

We discuss the implications that new magnetocaloric, thermal expansion and magnetostriction data in $\alpha$-RuCl$_{3}$ single crystals have on its temperature-field phase diagram and uncover the magnetic-field dependence of an apparent energy gap structure $\Delta (H)$ that evolves when the low-temperature antiferromagnetic order is suppressed. We show that, depending on how the thermal expansion data are modeled, $\Delta (H)$ can show a cubic field dependence and remain finite at zero field, consistent with the pure Kitaev model hosting itinerant Majorana fermions and localized $\mathbb{Z}_{2}$ fluxes. Our magnetocaloric effect data provides, below $1\,\mathrm{K}$,  unambiguous evidence for dissipative phenomena at $H_{\mathrm{c}}$, a smoking gun for a first-order phase transition. Conversely, our results show little support for a phase transition from a QSL to a polarized paramagnetic state above $H_{\mathrm{c}}$.

\end{abstract}

\maketitle

\section{Introduction}

The Kitaev model treats $S=1/2$ spins on a honeycomb lattice with bond-dependent interactions and is one of the few examples of an exactly solvable quantum spin-model on a two-dimensional (2D) lattice.  It has been shown that its ground state is a quantum spin liquid (QSL) with Majorana fermions and $\mathbb{Z}_{2}$ fluxes as fundamental excitations \cite{kitaev_anyons_2006, kitaev_fault-tolerant_2003}. The proposal \cite{jackeli_mott_2009} that Mott insulators with strong spin orbit coupling and the correct geometry could display Kitaev interactions and the associated QSL ground state stimulated significant research into candidate materials. Initially these focused on the iridate compounds Li$_{2}$IrO$_{3}$ and Na$_{2}$IrO$_{3}$, but more recently significant attention has been paid to $\alpha$-RuCl$_{3}$ \cite{chaloupka_magnetic_2016, koitzsch_j_2016, plumb__2014, kim_kitaev_2015} and monolayer of chromium compounds such as CrSiTe$_3$ \cite{xu_interplay_2018, xu_possible_2020}. The effective magnetic Hamiltonian for the materials includes Kitaev terms that may be anisotropic, as well as Heisenberg and off-diagonal exchange terms. In the absence of a magnetic field, stoichiometric $\alpha$-RuCl$_{3}$ single crystals show a sharp transition to antiferromagnetic (AFM) order around $T_{\mathrm{N}}=7\,\mathrm{K}$. The RuCl$_{3}$ layers are weakly coupled by van der Waals forces, and thus stacking faults are easily formed. These faults along with other defects or disorder lead to additional transitions, most prominently a broad transition around $14\,\mathrm{K}$ related to a different AFM stacking order \cite{cao_low-temperature_2016}.  The AFM order in $\alpha$-RuCl$_{3}$ can be readily suppressed by the application of a magnetic field; however, a definitive determination of the temperature-field phase diagram has proved so far elusive.

Indeed, numerous studies have recently proposed a field-induced QSL phase in $\alpha$-RuCl$_{3}$, based on the observation of unusual physical properties \cite{kubota_successive_2015, majumder_anisotropic_2015}, such as the emergence of a plateau in the thermal Hall effect \cite{kasahara_majorana_2018, yokoi_half-integer_2020}, the opening of a spin gap from thermal conductivity \cite{hentrich_unusual_2018}, specific heat \cite{wolter_field-induced_2017}, electron spin resonance (ESR) \cite{ponomaryov_unconventional_2017} and nuclear magnetic resonance (NMR) \citep{baek_evidence_2017, zheng_gapless_2017, nagai_two-step_2020, jansa_observation_2018} measurements. There are two types of quasiparticles arising from the fractionalization of the spin degree of freedom in a QSL: Majorana fermions and $\mathbb{Z}_{2}$ fluxes. Majorana fermions are itinerant, charge-neutral spin-1/2 particles that are their own antiparticles. They are excited with a gapless continuum, whereas the localized $\mathbb{Z}_{2}$ fluxes are gapped \cite{kitaev_anyons_2006}. In a magnetic field Majorana fermions also acquire a gap with a cubic field dependence, whereas the $\mathbb{Z}_{2}$ excitations are insensitive to the magnetic field \cite{kitaev_anyons_2006}. As pointed out by Nagai \textit{et al.} \cite{nagai_two-step_2020} the field dependence of the spin-gap is, however, controversial in the literature with specific heat and NMR studies reporting a vanishing gap around the critical field $\mu_{0}H_{\mathrm{c}} \simeq 7\,\mathrm{T}$ as well as a scaling behavior that is in agreement with a quantum critical point at $H_{\mathrm{c}}$ \cite{baek_evidence_2017, wolter_field-induced_2017, zheng_gapless_2017}, while other experiments indicate the presence of a finite residual gap in zero field \cite{jansa_observation_2018, ponomaryov_unconventional_2017, banerjee_excitations_2018} or a two energy-gap structure \cite{nagai_two-step_2020}.

Here we report in-plane lattice effects that underscore the strong spin-lattice coupling in $\alpha$-RuCl$_{3}$ single crystals, in both the AFM order and the field-induced states. This is complemented with magnetocaloric effect (MCE) measurements performed in pulsed magnetic fields at temperatures down to $0.56\,\mathrm{K}$. Earlier dilatometry work \cite{he_uniaxial_2018, gass_field-induced_2020} focused mainly on the lattice change along the out-of-plane crystallographic direction [0,0,1]. By contrast, we perform thermal expansion and magnetostriction measurements directly probing the change of the relevant in-plane lattice parameters along [1,1,0] using a fiber Bragg grating method (FBG) \cite{daou_high_2010, jaime_fiber_2017}. These results reveal an energy gap, \textit{i.e.} an energy scale, that, depending on the modeling, does not vanish at $H_{\mathrm{c}}$. Magnetocaloric effect measurements in a $^3$He refrigerator down to $0.5\,\mathrm{K}$ were accomplished in two limits, quasi-adiabatic and quasi-isothermal \cite{kohama_acsh_2010, kihara_mce_2013}, that show previously undetected dissipative mechanisms closely related to the suppression of AFM order at $H_{\mathrm{c}}$.  

\begin{figure}
\includegraphics{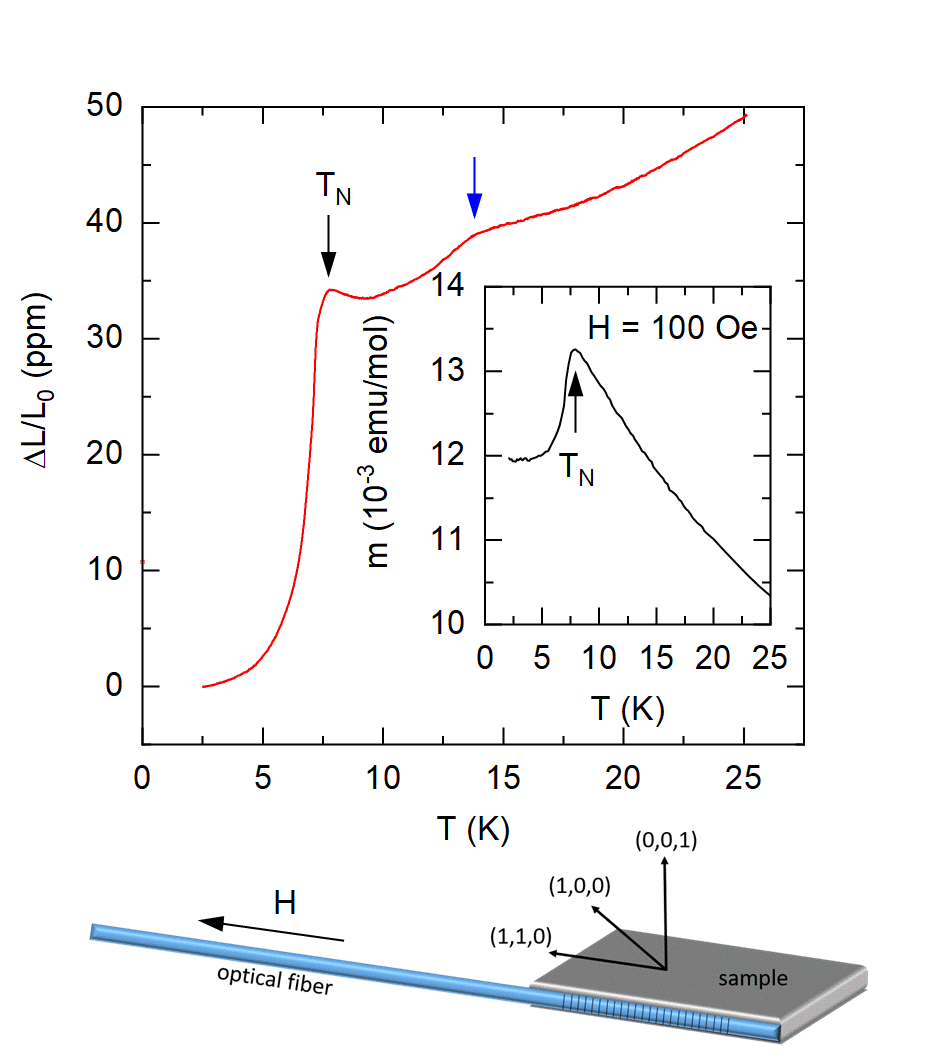}
\caption{Thermal expansion $\Delta L/L_{0}$ as a function of temperature. The arrows indicate the onset of the antiferromagnetic ordering temperature $T_{\mathrm{N}}$ at around $7\,\mathrm{K}$ (black arrow) as well as the broad feature present at $14\,\mathrm{K}$ (blue arrow). The inset shows the low-temperature magnetization of the same crystal recorded at a magnetic field of $100\,\mathrm{Oe}$. A depiction of the fiber Bragg grating setup is shown in the bottom. The magnetic field is aligned along the optical fiber and parallel to the [1,1,0] crystallographic direction. }
\label{fig:figure1}
\end{figure}

\section{Results}

Single crystal samples of $\alpha$-RuCl$_{3}$ were prepared using high-temperature vapor transport techniques from pure $\alpha$-RuCl$_{3}$ powder with no additional transport agent\cite{banerjee_neutron_2017}. All crystals reported here exhibit a single dominant transition temperature of $T_{\mathrm{N}}\simeq 7\,\mathrm{K}$ comparable to other recent studies \cite{cao_low-temperature_2016, balz_finite_2019, sears_magnetic_2015}, indicative of high-quality crystals with minimal stacking faults. Characterization carried out by means of magnetization measurements on the $\alpha$-RuCl$_{3}$ single crystals in a Quantum Design magnetic properties measurement system (MPMS) is consistent with previous studies \cite{plumb__2014, zheng_gapless_2017, wang_magnetic_2017}. 

\begin{figure*}
\includegraphics{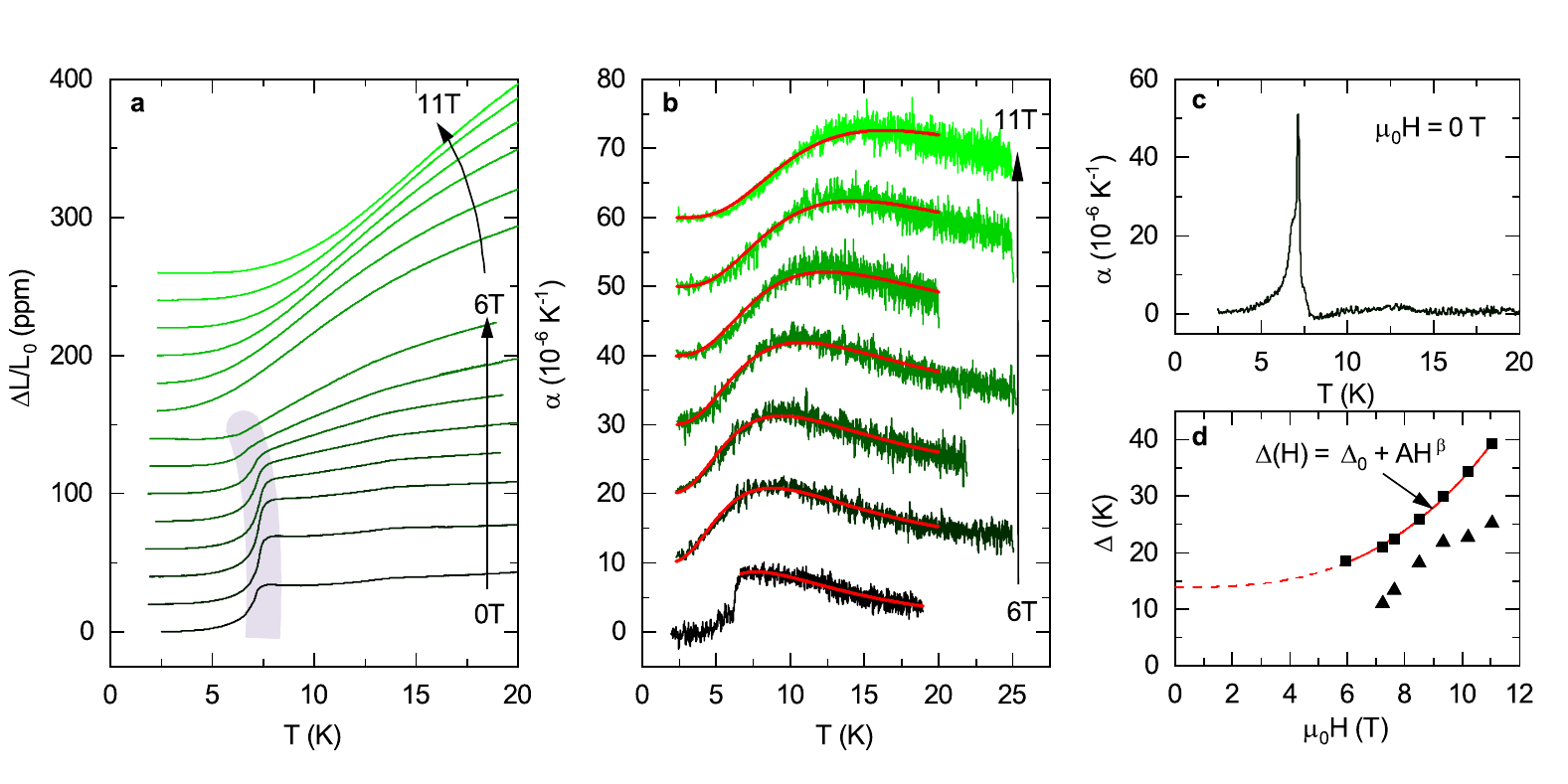}
\caption{(a) Thermal expansion $\Delta L/L_{0}$ vs. temperature for different magnetic fields. Curves were shifted vertically for clarity. The gray shaded region indicates the antiferromagnetic transitions. (b) Thermal expansion coefficient $\alpha $ for fields $H > 6\,\mathrm{T}$. The solid red lines are fits representing a Schottky-like behavior according to equation \ref{eq:equation1} while the curve recorded at $6\,\mathrm{T}$ was fitted only for temperatures above $7\,\mathrm{K}$, avoiding  the antiferromagnetic transition. (c) Thermal expansion coefficient $\alpha$ vs. temperature at zero magnetic field. (d) The gap $\Delta$(H) (squares), indicating a $H^{3}$ behavior (solid red line) and a finite gap $\Delta_{0}$ at $\mu_0H = 0\,\mathrm{T}$. The triangles represents the activation energy extracted from fits using a single exponential function.}
\label{fig:figure2}
\end{figure*}

\subsection{Thermal expansion}

Figure \ref{fig:figure1} shows the thermal expansion vs. temperature of $\alpha$-RuCl$_{3}$. The optical fiber was attached to the side of the sample parallel to the [1,1,0] crystallographic direction, thus probing the change of the in-plane lattice parameters. A sharp drop in $\Delta L/L_{0} = (L(T)-L(T_{\mathrm{min}}))/L_{0}$, where $L_{0}$ is the length of the fiber Bragg grating ($5\,\mathrm{mm}$), can be observed at the AFM ordering temperature $T_{\mathrm{N}}$ around $7\,\mathrm{K}$. The broad feature present around $14\,\mathrm{K}$ is attributed to the presence of stacking faults \cite{cao_low-temperature_2016}. The magnetization data for the same crystal in Fig. \ref{fig:figure1} inset shows a single step like transition at $T_{\mathrm{N}}$ and no feature at higher temperatures. However, magnetization can be a less sensitive probe than specific heat or thermal expansion in detecting the $14\,\mathrm{K}$ transition.  Some stacking faults might be induced by the cutting process, which is unavoidable for this experiment. The FBG method is potentially more sensitive to stacking faults in the vicinity of the cut edge than the magnetization, which probes the entire sample. A small amount of residual strain on the sample, caused by the differential thermal contraction between sample and optical fiber, cannot be ruled out as well. 

The magnetic field dependence of $T_{\mathrm{N}}$ can be tracked by thermal expansion measurements as shown in Fig. \ref{fig:figure2}(a). The AFM transition is visible as a drop in the relative change of length $\Delta L/L_{0}$ signal at $T_{\mathrm{N}}$(H) in applied fields up to $6\,\mathrm{T}$, as well as a peak in the coefficient of linear thermal expansion $\alpha(T) = (1/L_{0})(\partial\Delta L/\partial T)$ shown in Fig. \ref{fig:figure2}(c). Remarkably $\alpha(T)$ obtained at  $H \geq 6\,\mathrm{T}$ show a broad feature resembling a Schottky-type anomaly indicating the presence of an energy gap [see Fig. \ref{fig:figure2}(b)]. We, therefore, fit  the temperature dependence of the thermal expansion coefficient with the equation:

\begin{equation}
\alpha(T) \approx \frac{R\Delta^{2}}{T^{2}}\frac{\mathrm{e}^{\Delta /T}}{(1+\mathrm{e}^{\Delta /T})^{2}},
\label{eq:equation1}
\end{equation}

\noindent where $R$ is a constant and $\Delta$ an energy gap. The presence of a Schottky-type anomaly in the thermal expansion data is a somewhat common phenomenon since heat capacity and thermal expansion coefficient are both second derivatives of the free energy. The effect has been observed before primarily in $f$-electron compounds where the energy gap is associated with the relatively small crystal field splitting (see, for example, Refs. \cite{ott_crystal-electric-field_1976, novikov_heat_2012}). In $\alpha$-RuCl $\Delta$ increases as a function of the applied field as depicted in Fig. \ref{fig:figure2}(d). Furthermore, $\Delta(H)$ can be described by a single power law $\Delta(H) = \Delta_{0} + aH^{\beta}$ where $\Delta_{0}$ is the gap value at $H = 0$. The zero field gap $\Delta_{0}$ and the exponent $\beta$ can be extracted from the fit: One obtains $\Delta_{0} = (14\pm0.1)\,\mathrm{K}$ and $\beta = 2.9\pm 0.1$. Figure 1 in the supplementary material illustrates the robustness of the fitting parameters \cite{note_supplement}. The nearly cubic field dependence of $\Delta (H)$ is consistent with a Majorana fermion gap \cite{kitaev_anyons_2006} as seen before in NMR measurements \cite{jansa_observation_2018, nagai_two-step_2020}. The finite zero-field gap $\Delta_{0}$ can be associated with the field-independent $\mathbb{Z}_{2}$ flux.

Indeed, for magnetic fields below $6\,\mathrm{T}$ no Schottky anomaly can be identified, likely due to the presence of the AFM transition which dominates the temperature dependence of the thermal expansion coefficient [see Fig. \ref{fig:figure2}(c)]. Note that the $6\,\mathrm{T}$ curve in Fig. \ref{fig:figure2}(b) already shows a significant drop in $\alpha$ due to the magnetic ordering, however it is still possible to fit the high temperature tail to Eq. (\ref{eq:equation1}) and extract the gap value as depicted in the graph. Similarly to earlier specific heat measurements \cite{sears_phase_2017} we also fit the low temperature tails of $\alpha (T)$ with a single exponential activation function $\alpha (T) = Ae^{-\frac{\Delta}{T}}$ (see Fig. 2 in the supplementary material \cite{note_supplement}) which yields a field dependent gap [triangles in Fig. \ref{fig:figure2}(d)] that is comparable with those extracted from specific heat and thermal conductivity measurements \cite{sears_phase_2017, wolter_field-induced_2017, hentrich_unusual_2018}.

\begin{figure}
\includegraphics{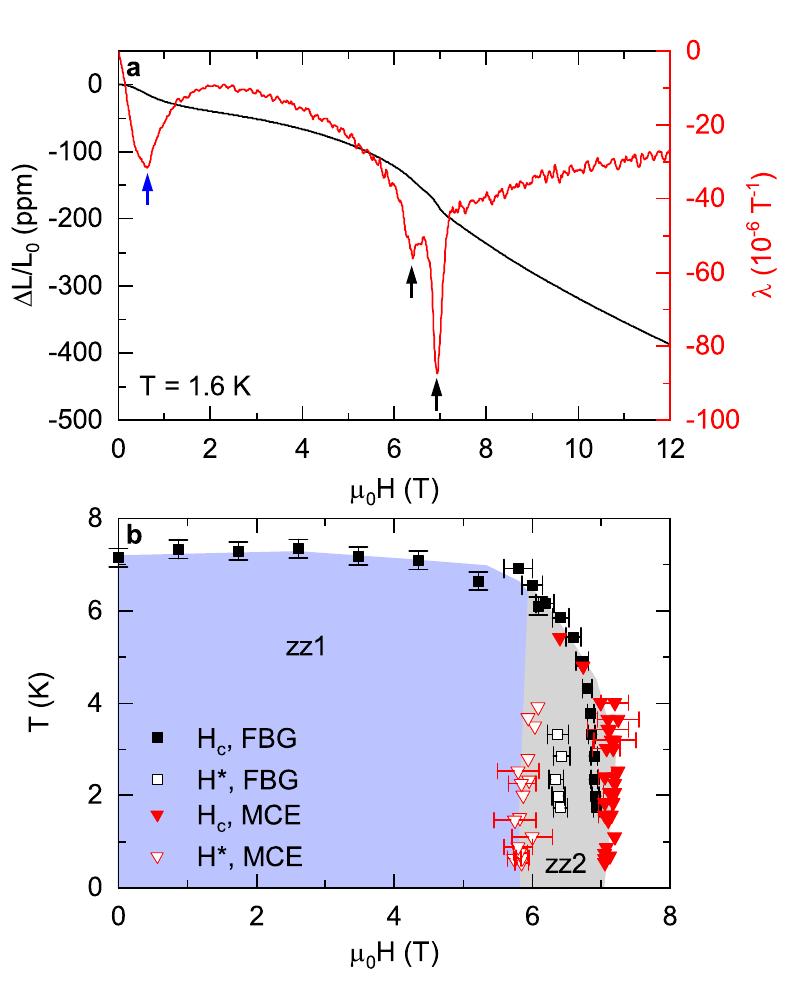}
\caption{(a) $\Delta L/L$ \textit{vs} the magnetic field at $T = 1.6\,\mathrm{K}$ (black curve). The arrows denote the critical fields marking anomalies in the derivative $\lambda = (1/L_{0})\partial(\Delta L/L)/\partial H$ (red line) between the different AFM phases zz$_1$ and zz$_2$, as well as AFM and the partially field polarized paramagnetic state. The low field anomaly (blue arrow) is likely caused by magnetic domain flip as mentioned in the text. (b) Phase diagram of $\alpha$-RuCl$_{3}$ extracted from magnetoelastic and MCE measurements. Open symbols represent the transition from the zz$_1$ to the zz$_2$ AFM state, and solid symbols indicate the transition between AFM order and the partially field polarized PM state.}
\label{fig:figure3}
\end{figure}

\subsection{Magnetostriction}

Under an external magnetic field $\alpha$-RuCl$_{3}$ shows lattice contraction along [1,1,0] as depicted in Fig. \ref{fig:figure3}(a). The critical field $H_{\mathrm{c}}$, which leads to the destruction of the AFM order and the onset of a partially field polarized magnetic phase, manifests as a kink in the $\Delta L/L_{0} $ vs. H  data, where $\Delta L = L(H) - L(H=0)$. A second transition at a lower field $H^{\ast}$ is evident in the linear magnetostriction coefficient $\lambda = (1/L_{0})\partial\Delta L/\partial H$ and corresponds to the transition between two different AFM phases with different stacking order. These phases have been identified in the literature \cite{balz_finite_2019, lampen-kelley_field-induced_2018} and labeled zz$_1$ and zz$_2$. The low field anomaly in $\alpha(H)$ around $1\,\mathrm{T}$ is likely a redistribution of magnetic domain populations \cite{sears_phase_2017}. It is interesting to note that the significant lattice parameter reduction in the high magnetic field state is, by itself and independently of the specific functional dependence chosen to fit the data, unambiguous evidence for a finite energy gap. The magnetostriction curves are almost temperature independent to $T = 7\,\mathrm{K}$ and $\mu_{0}H = 10\,\mathrm{T}$ aside from the above mentioned features that indicate the phase transitions. The resulting AFM phase diagram of $\alpha$-RuCl$_{3}$ is shown in Fig. \ref{fig:figure3}(b) containing both the FBG and MCE data discussed below. The critical fields were extracted from minima in $\lambda (H)$. Our phase diagram and critical fields agree well with the literature \cite{balz_finite_2019, lampen-kelley_field-induced_2018}.

\subsection{Magnetocaloric effect}

\begin{figure}[ht]
\includegraphics{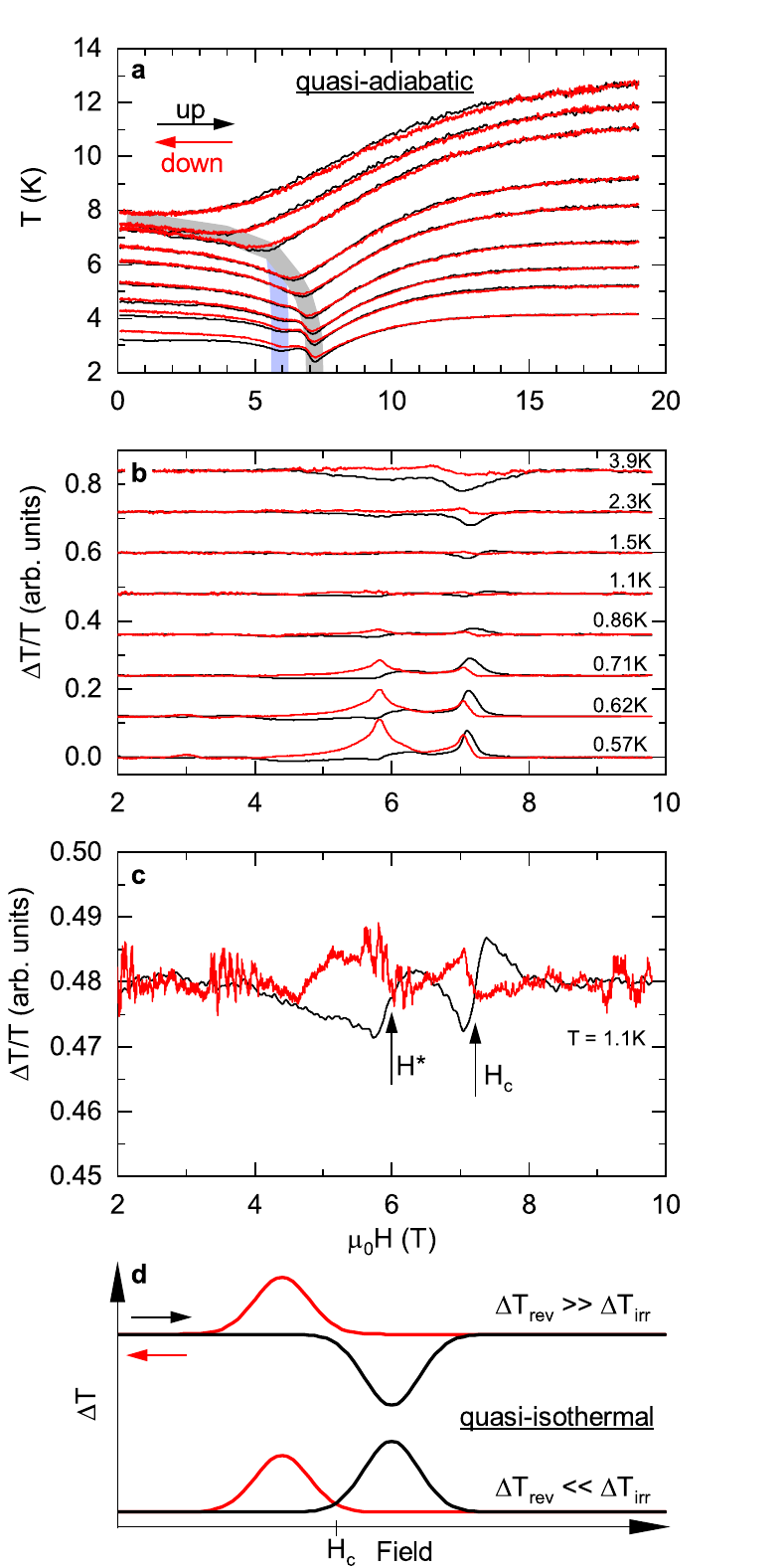}
\caption{(a) Sample temperature vs. magnetic field under quasi-adiabatic conditions. The shaded areas mark the first (blue) and second (gray) critical field respectively. The arrows symbolize the field sweep direction. (b) Relative change of the sample temperature $\Delta T/T$ as a function of the magnetic field under quasi-isothermal conditions, after a smooth background subtraction. (c) Expanded MCE curve at an initial temperature of $1.1\,\mathrm{K}$. (d) Cartoon displaying the expected quasi-isothermal behavior of the MCE under reversible (second-order) and irreversible (first-order) conditions\cite{silhanek_irreversible_2006}.}
\label{fig:figure4}
\end{figure}

Results from our MCE measurements are discussed in two parts. First, measurements under quasi-adiabatic conditions were carried out at temperatures above $2\,\mathrm{K}$ by removing the ${}^{4}$He exchange gas, which decreases the thermal conductivity between the sample and the bath. For these measurements the magnetic field was also applied along the [1,1,0] axis. The resulting temperature-field curves shown in Fig. \ref{fig:figure4}(a) display an almost perfectly reversible behavior on sweeping the magnetic field up and down above $7\,\mathrm{K}$, confirming quasi-adiabatic conditions, i.e., virtually no heat exchange with the thermal bath. The open loops below $7\,\mathrm{T}$, seen as separation between up and down field curves, are attributable to some irreversible heating at the phase boundaries characteristic of first-order-like phase transitions.

The critical fields $H^{\ast}$ and $H_{\mathrm{c}}$ manifest as minima and/or shoulders in the temperature vs. $H$ curves and were added to the phase diagram shown in Fig. \ref{fig:figure3}(b). The temperature drop at critical fields is typical due to the increased magnetic entropy in the proximity of phase transitions. The zz$_2$ phase appears to be much narrower in field in the dilatometry data, when compared to the MCE effect. However, so far we can only speculate about the possible reasons for this behavior and the difference might also be related to the criteria chosen to define $H^{\ast}$ and $H_{\mathrm{c}}$. A small in-plane sample misalignment ($\approx 5^{\circ}$) between the MCE and FBG measurements can lead to a shift in the transition fields and a narrower zz$_{2}$ phase \cite{lampen-kelley_field-induced_2018}. Additionally, strain induced by the thermal expansion mismatch between the optical fiber and the $\alpha$-RuCl$_{3}$ sample might also lead to different critical fields, since the MCE sample is not attached to any substrate or fiber and thus strain free. The critical field values shown for this type of measurement are in excellent agreement with earlier reports \citep{balz_finite_2019, lampen-kelley_field-induced_2018}.

Adiabatic conditions at temperatures below $2\,\mathrm{K}$ are difficult to realize due to residual liquid ${}^{4}$He in the sample space and adsorbed ${}^4$He atoms on the sample surface, as well as the lack of sufficient cooling power in the absence of residual liquid. In order to study the phase diagram at temperatures below $2\,\mathrm{K}$ we conducted MCE measurements under quasi-isothermal conditions (also called ``equilibrium`` in Ref. \cite{zapf_bose-einstein_2014}) where the sample was immersed in either liquid ${}^{3}$He or ${}^{4}$He ensuring a good thermal link between the sample and bath. This finite (good enough to cool down the sample yet far from perfect) thermal link results in a finite temperature change $\Delta T$ as a function of the magnetic field close to the critical fields and a otherwise constant sample temperature. The scaled $\Delta T$ vs. $H$ curves, after subtraction of a smooth background, are shown in Fig. \ref{fig:figure4}(b) and (c). Low temperature features are enhanced compared to the higher-temperature data (e.g., the $4\,\mathrm{K}$ curve) as a result of the temperature scaling, the raw data are included in the supplemental material (Fig. 3) \cite{note_supplement}. 

At temperatures of $\approx 1\,\mathrm{K}$ and above our data agrees well with recent results by Balz \textit{et al.} \citep{balz_finite_2019}, consistent with a continuous (second-order-like) transition from the AFM state. Due to magnetic fluctuations and an increased entropy, a reduction of the sample temperature is observed in the quasi-adiabatic data as the magnetic field approaches $H^{\ast}$. After passing the phase transition the sample temperature slowly relaxes back towards the bath temperature and further cooling is observed by crossing the phase boundary at $H_{\mathrm{c}}$. Consequently a positive temperature change $\Delta T$ is observed when entering the AFM phase by decreasing the magnetic field. Under quasi-isothermal conditions, the magnitude $\Delta T$ is not the same during the up- and down-sweep due to the different sweep rates and irreversible contributions. Similarly to Ref. \cite{silhanek_irreversible_2006} one can split $\Delta T$ into an reversible and irreversible component $\Delta T = \Delta T_{\mathrm{rev}} + \Delta T_{\mathrm{irr}}$. The irreversible temperature difference $\Delta T_{\mathrm{irr}}$ represents dissipative processes inside the sample which always lead to an increase in the sample temperature, regardless in which direction the phase boundary is crossed.

Remarkably, the MCE curves clearly show irreversible behavior below $1\,\mathrm{K}$, meaning that while going through the AFM phase transition in both up- and down-sweep the sample temperature increases, indicating the release of latent heat or other irreversible processes such as AFM domain movement. The transition at $H^{\ast}$ is expected to be first-order due to the coexistence of phases with a different AFM ordering vector. The second phase transition at $H_{\mathrm{c}}$ also shows a large irreversible component at low temperatures. Due to the vanishing entropy and $T \rightarrow 0$, $\Delta T_{\mathrm{rev}}$ becomes smaller at lower temperatures and $\Delta T_{\mathrm{irr}}$ dominates below $1\,\mathrm{K}$. This strongly indicates a first-order phase transition below $1\,\mathrm{K}$ as opposed to a second-order phase transition which is characterized by a reversible temperature behavior displaying sample cooling during the up-sweep and heating during the down-sweep [see Fig. \ref{fig:figure4}(d)].Whether the transition remains first order at higher temperatures under quasi-adiabatic conditions cannot be clearly determined from our data partially due to the experimental resolution and overlap with the transition at $H^{\ast}$.

\section{Discussion}

The first point we want to address in our discussion is the presence of the energy gap feature which emerges from the thermal expansion measurements. Due to the connection between $\alpha$ and $c_{\mathrm{p}}$ it is natural to assume that $\Delta$ can be associated with the spin gap observed in previous specific heat studies of $\alpha$RuCl$_{3}$ even though it is difficult to extract the purely magnetic contribution to $\alpha (T)$, one can assume that the nonmagnetic (phonon) contribution to $\alpha (T)$ is field independent. Indeed, the analysis of the low-temperature activation behavior above $H_{\mathrm{c}}$ yields a similar gap size and $\Delta(H)$ behavior when compared to the specific heat and thermal conductivity measurements \cite{sears_phase_2017, wolter_field-induced_2017, hentrich_unusual_2018}, thus indicating that the exact field dependence of the spin gap strongly depends on which model and temperature range is chosen to extract the gap size. Our approach yields two energy gaps that overall are consistent with the two-gap structure reported in Ref. \cite{nagai_two-step_2020}. The gap size and the increasing field dependence are also in agreement with recently reported ESR and neutron scattering data \cite{ponomaryov_nature_2020, balz_finite_2019}. Note that in those studies a reopening energy gap is observed for fields below $H_{\mathrm{c}}$ that are not captured by our measurements.

Second, we discuss our results in the light of a potential field-induced QSL phase or proximate QSL behavior above $H_{\mathrm{c}}$. In the past, significant efforts have been made to reveal the field-induced quantum spin liquid phase in $\alpha$-RuCl$_{3}$. Most prominently, studies of the thermal hall effect \cite{kasahara_majorana_2018, yokoi_half-integer_2020} report evidence for fractional excitations based on the emergence of a half integer plateau within a finite field range between approximately 10 and $11\,\mathrm{T}$ for $H\parallel [1,1,0]$ \cite{yokoi_half-integer_2020}. This suggests the presence of additional phased transitions above $H_{\mathrm{c}}$ between the quantum spin liquid and polarized paramagnetic state. Signs for transitions above $H_{\mathrm{c}}$ were detected by MCE measurements at around $9\,\mathrm{T}$ \cite{balz_finite_2019} and magnetostriction experiments at approximately $11\,\mathrm{T}$ \cite{gass_field-induced_2020}. Other studies reporting measurements of the magnetic Gr{\"u}neisen parameter and specific heat \cite{bachus_thermodynamic_2020} as well as ESR \cite{ponomaryov_nature_2020} show no signatures of phase transition beyond $8\,\mathrm{T}$.

The magnetoelastic and magnetocaloric data presented here show no evidence for a field-induced phase transition above $H_{\mathrm{c}}$ even at temperatures as low as $0.57\,\mathrm{K}$. Only a broad maximum in the isothermal MCE is visible around $12.5\,\mathrm{T}$ (see supplementary material Fig. 3 \cite{note_supplement}) not indicative of a phase transition.

Note that in this work the magnetic field is aligned perpendicular to the Ru-Ru bonds, whereas in Refs. \cite{balz_finite_2019, gass_field-induced_2020, bachus_thermodynamic_2020} different in-plane field orientations were chosen. This calls for further studies to evaluate whether additional high field phase transitions are present for different in-plane field orientations.

\section{Summary}

In summary, we conducted new measurements of the lattice and thermal properties of $\alpha$-RuCl$_{3}$. We observe an energy gap which follows a $H^{3}$ behavior. It is, however, unclear whether this behavior can me assigned to fractional excitations as in recent specific heat measurements \cite{tanaka_thermodynamic_2020} or if it can be attributed to conventional magnons \cite{ponomaryov_nature_2020}. Thermal measurements for fields applied perpendicular to the Ru-Ru bonds show clear evidence for a first-order phase transitions at $H^{\ast}$ and $H_{c}$. No signature of a transition between the proposed field-induced quantum spin liquid and the high field paramagnetic state was found. These results place strong constraints on any theory put together to explain quantum critical behavior and the phenomenology of a QSL phase in $\alpha$-RuCl$_{3}$.

\section*{Acknowledgements}

We thank M. B. Salamon, C. D. Batista, Y. Kohama, M. Lee, and V. Zapf for helpful discussions. A portion of this work was performed at the National High Magnetic Field Laboratory, which is supported by the NSF Cooperative Agreement No. DMR-1644779, the U.S. DOE and the State of Florida. This material is based upon work supported by the U.S. Department of Energy, Office of Science, National Quantum Information Science Research Centers. P.F.S.R. acknowledges support from the Los Alamos Laboratory Directed Research and Development program through project 20210064DR. D.G.M. acknowledges support from the Gordon and Betty Moore Foundation’s EPiQS Initiative, Grant No. GBMF9069. J.Q.Y. acknowledges support from the U.S. Department of Energy, Office of Science, Basic Energy Sciences, Materials Sciences and Engineering Division. S.N. was supported by the U.S. DOE Office of Science, Basic Energy Sciences, Division of Scientific User Facilities, and R.S., Y.T. and M.J. by the NHMFL UCGP program.

\bibliography{RuCl3_MCE}

\end{document}


\title{Supplemental Material}

\maketitle

\section{FBG Dilatometry}

FBG dilatometry measurements were performed in a superconducting magnet system equipped with a ${}^{4}$He cryostat. The plate-like sample was carefully cut using a wire saw to provide a straight edge where the optical fiber is attached to, using a cyano-acrylate based adhesive. To ensure a good thermal anchoring, a gold wire was glued to the sample, providing a thermal link to the bath. The FBG spectra were recorded with an optical sensing interrogator (Micron Optics, si155)\textsuperscript{\textregistered}. A second and third Bragg grating present in the optical fiber were used as references. The reference thermal expansion and magnetostriction signals were fitted and subsequently subtracted from the sample signal. The temperature dependence of the refractive index and the thermal expansion coefficient of the SiO$_{2}$ fiber are small in the measured temperature range ($T<40\,\mathrm{K}$) and can be neglected (see M. Jaime, C. et al., Sensors 17, 2572 (2017)).

\section{Magnetocaloric Effect}

For the MCE measurements we deposited a thin film $\approx10\,\mathrm{nm}$ of AuGe (Au $16\%$ target) on the sample surface which acts as a thermometer due to its semiconducting resistivity behavior. This method ensures an excellent thermal link between the sample and the thermometer which is crucial given the short millisecond-long timescales in pulse field experiments. To improve the contact resistance a thin layer of Gold $\approx 10\,\mathrm{nm}$ was deposited on top of the AuGe; part of the sample was masked to avoid shorting the AuGe film. The resistance of the AuGe thermometer was then measured via a quasi two contact measurement. Typical resistances of the AuGe film reach around $400\,\Omega$ at $4\,\mathrm{K}$ - far above any contact resistances present during this measurements. The MCE measurements were performed in a $65\,\mathrm{T}$ short pulse magnet at the National High Magnetic Field Laboratory pulsed field facility at Los Alamos National Laboratory. The magnetic field pulse has an approximate rise time of $5\,\mathrm{ms}$ and a total pulse length of $\approx 30\,\mathrm{ms}$. The sample was not glued to the sample holder to limit the thermal anchoring for adiabatic measurement conditions. Quasi-adiabatic conditions were achieved by cooling down the sample and removing all cryogenic liquid and exchange gas surrounding the sample. On the other hand, under quasi-isothermal conditions the sample was immersed in liquid ${}^3$He or ${}^4$He, ensuring a better connection to the thermal bath. An AuGe film deposited on the glass slide was used as a reference to subtract the magnetoresistance of the film. The sample temperature was calibrated with a nearby calibrated cernox thermometer.

(i) Quasi-adiabatic conditions imply that the thermal link between the sample and the thermal bath is weak and it can be ideally assumed that there is no heat exchange between the sample and bath during the field pulse. Thus, the sample temperature should behave completely reversible with respect to the field up and down sweep assuming that no dissipative processes occur during the field sweep. Experimentally we cannot reach perfect adiabatic conditions since residual He gas or liquid as well as the attached wires create a finite thermal link between bath and sample. In our case we limited the size of the thermal link by pumping on the sample space after cooldown and choosing thin constantan wire to curb the thermal conductivity between sample and bath. It is experimentally challenging to ensure quasi-adiabatic conditions especially in the temperature range $0.5$-$2.0\,\mathrm{K}$ that require liquid ${}^{3}\mathrm{He}$ - therefore we performed measurements in this temperature range under quasi-isothermal conditions. 

(ii) MCE measurements under quasi-isothermal conditions were performed in liquid ${}^{4}\mathrm{He}$ and ${}^{3}\mathrm{He}$ ensuring a better link between the sample and thermal bath. The sample temperature ideally only deviates from the bath temperatures in field regions where the sample experiences a large entropy change (e. g. phase transitions) and the sample does not have sufficient time to relax to thermal equilibrium with the bath. Under real experimental conditions we observe sample heating due to the large $\mathrm{d}B/\mathrm{d}t$ at the beginning of the pulse (see figure \ref{fig3}(b)), likely caused by improved thermal coupling between the sample and other metallic parts of the probe and cryostat. The effect is most recognizable at the base temperature around $0.6\,\mathrm{K}$. The sample temperature then relaxes slowly back to the inital value during the field up-sweep - the sample heating due to the first order phase transitions is superimposed onto this relaxation background. The background itself was approximated by a third order polynomial (red lines in figure \ref{fig3}(b)) in the field range between 2 and $10\,\mathrm{T}$ and subtracted from the data.

\section{Supplemental Figures}

\begin{figure}
\includegraphics{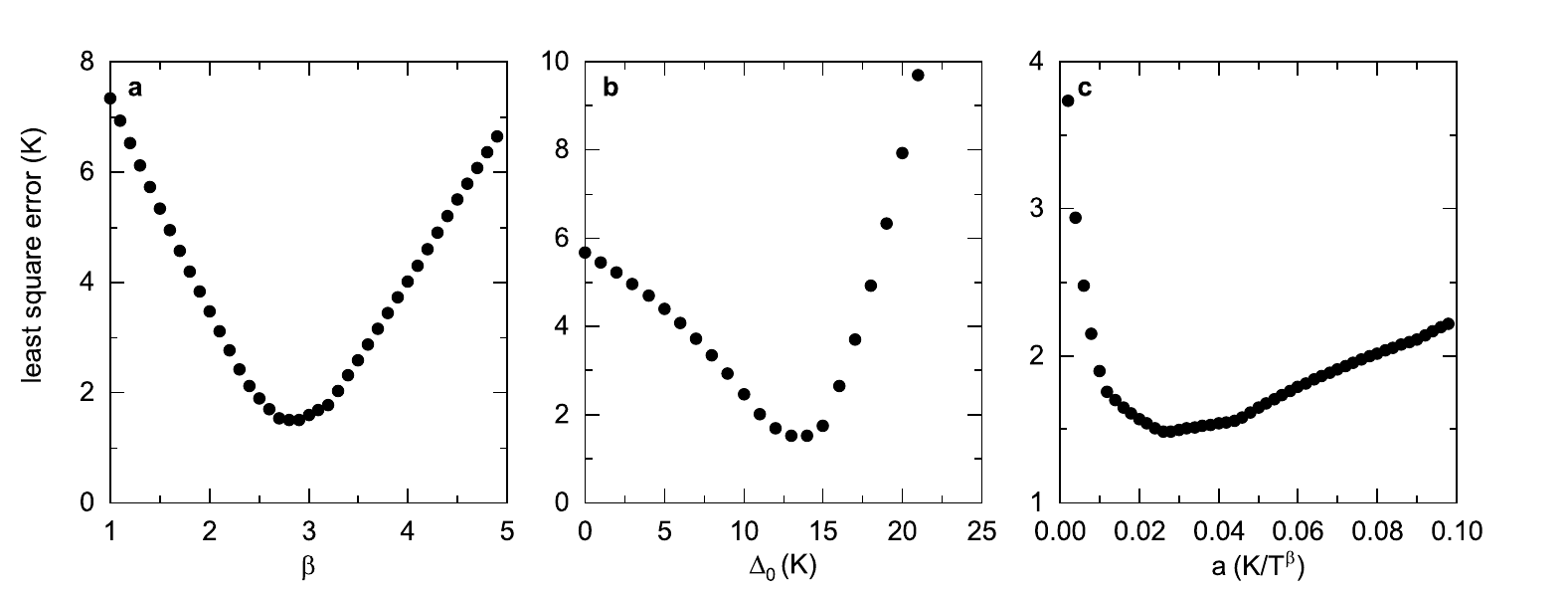}
\caption{(a, b, c) Least square error as a function of the fitting parameters $\beta$, $\Delta_{0}$ and $a$ used in the power law fit of the energy gap $\Delta = \Delta_{0} + a(\mu_{0}H)^{\beta}$. The variable on the abscissa is fixed in each plot and the remaining two parameters are fitted to minimize the error.}
\end{figure}

\begin{figure}
\includegraphics{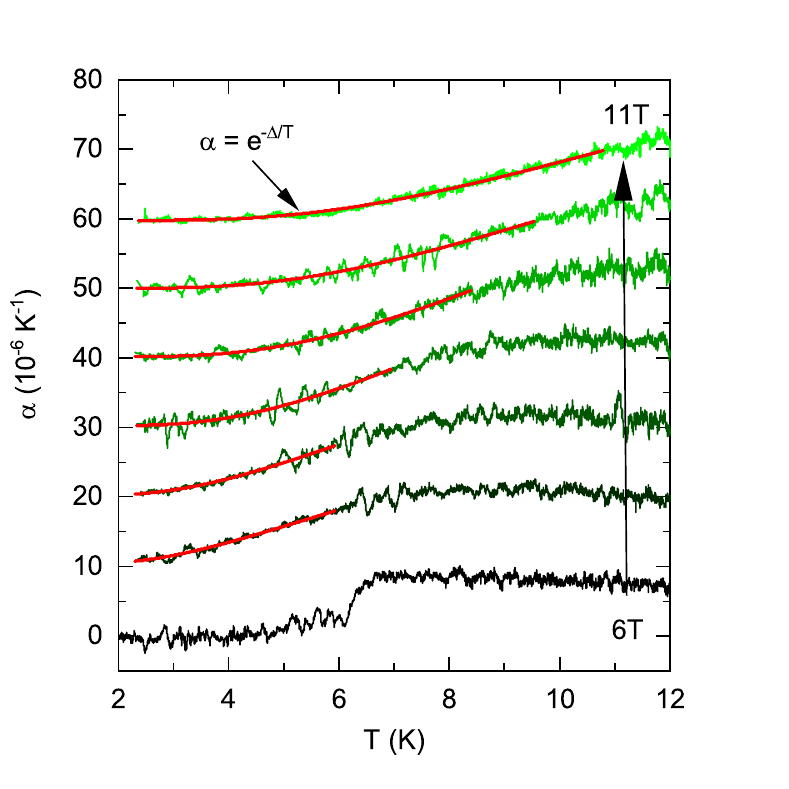}
\caption{Low temperature behavior of the thermal expansion coefficient $\alpha$ for magnetic fields above $6\,\mathrm{T}$. The curves are shifted vertically for clarity. The red lines represent fits of a single exponential activation function $\propto Ae^{-\frac{\Delta}{T}}$ of the low temperature part of $\alpha(T)$.}
\end{figure}

\begin{figure}
\includegraphics{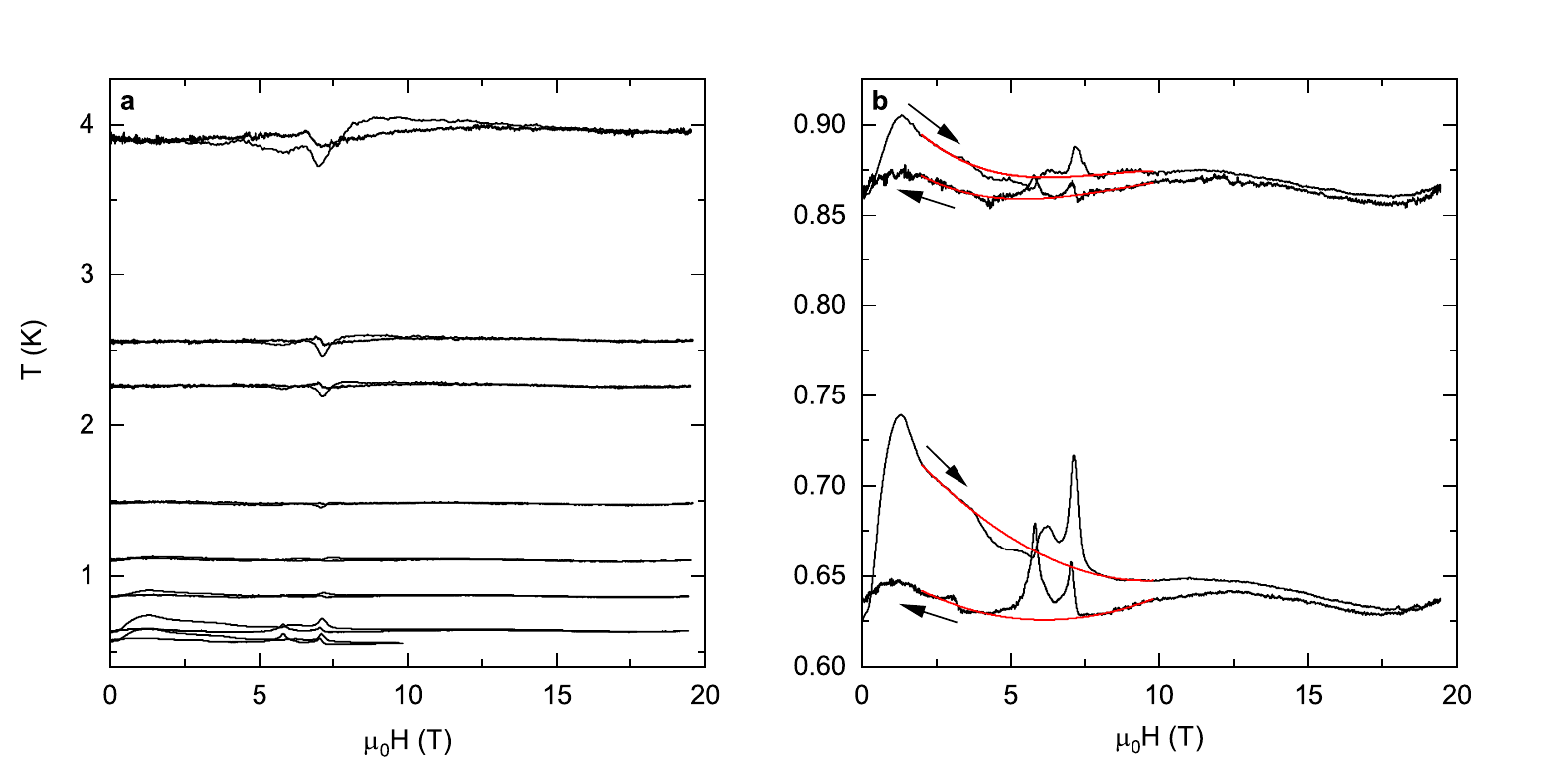}
\label{fig3}
\caption{(a) Magnetocaloric effect data, displaying the sample temperature as a function of the magnetic field. (b) At lower temperatures, due to the poorer thermal coupling to the bath and the large $\mathrm{d}H/\mathrm{d}t$ at the beginning of the pulse, a significant sample heating is observed. We accounted for this effect by fitting the MCE data with a third order polynomial (red lines) in the field range from 2 to $10\,\mathrm{T}$.}
\end{figure}
